\documentclass[useAMS,usenatbib]{mn2e}
\usepackage{graphicx}
\usepackage{epstopdf}

\title[AGN/starburst content in high-$z$ ULIRGs]
{The AGN/starburst content in high redshift ULIRGs}

\author[Y. Watabe et al.]
{Y. Watabe, $^{1}$\thanks{E-mail: watabe@arcetri.astro.it} 
G. Risaliti, $^{1,2}$
M. Salvati, $^{1}$
E. Nardini, $^{3}$
E. Sani, $^{3}$
A. Marconi$^{3}$
\\
$^{1}$INAF-Osservatorio Astrofisico di Arcetri, L.go E. Fermi 5, 50125 Firenze, Italy\\
$^{2}$Harvard-Smithonian Center for Astrophysics, 60 Garden St. Cambridge, MA 02138 USA\\
$^{3}$Dipartimento di Astronomia e Scienza dello Spazio, Universit$\grave{a}$ di Firenze, L.go E. Fermi 2, 50125 Firenze, Italy}

\begin{document}

\date{}

%\pagerange{\pageref{firstpage}--\pageref{lastpage}} \pubyear{2002}

\maketitle

\label{firstpage}

\begin{abstract}
We apply a simple model, tested on local ULIRGs, to disentangle the active galactic nucleus (AGN) and starburst contributions
in submillimiter and 24~$\mu$m-selected ULIRGs observed with the $Spitzer$-IRS spectrometer.
We quantitatively estimate the average AGN contribution to the stacked $6-8$~$\mu$m rest-frame spectra of these sources in different
luminosity and redshift ranges, and, under the assumption of similar infrared-to-bolometric ratios as in local ULIRGs,
the relative AGN/starburst contributions to the total infrared luminosity. 
Though the starburst component is always dominant in submillimeter-selected ULIRGs, we find a significant increase of the AGN contribution at redshift $z>2.3$ with respect to lower $z$ objects. Finally, 
we quantitatively confirm 
that the mid-infrared emission of 24~$\mu$m-selected ULIRGs is dominated by the AGN component, 
but the starburst component contributes significantly to the bolometric luminosity.
\end{abstract}


%%%%%%%%%% Section 1: Introduction %%%%%%%%%%

\section{Introduction}
Infrared number counts and the cosmic infrared background provide strong evidence that  Ultraluminous Infrared Galaxies 
\citep[ULIRGs; e.g.,][]{So84,SM96,Bl02,Lo06}, 
i.e. sources with a total infrared luminosity, $L_{8-1000 \mu \rm m}$, greater than $10^{12} L_{\odot}$, are a key component of the high redshift Universe 
\citep[e.g.,][]{Do01,Lag05,Le05}. 
ULIRGs at high redshift have been discovered using different selections, leading to distinct
populations.
In particular, observations performed with the Submillimeter Common User Bolometer Array \citep[SCUBA;][]{Ho99} on the James Clerk Maxwell Telescope \citep[JCMT; e.g., ][]{Sm97} discovered a new population of submillimeter galaxies (SMGs), distributed at $z=1-3$ 
(median redshift is 2.3) \citep[e.g.,][]{Ch05},
 and with estimated
total infrared luminosity in the ULIRG range. We refer to these galaxies as submm-ULIRGs.

Another population of high-$z$ infrared luminous galaxies has been discovered through 24~$\mu$m selection. For example, 
\citet{Ho05} and \citet{Br07} have surveyed the Bo{\" o}tes field of the NOAO Deep Wide-Field Survey \citep[NDWFS;][]{JD99} with the Multiband Imaging Photometer for {\it Spitzer} \citep[MIPS;][]{Ri04} at 24 $\mu$m. Also, \citet{Ya05} have reported the initial results from a {\it Spitzer} GO-1 program to obtain low-resolution, mid-infrared spectra in the {\it Spitzer} First Look Survey (FLS). These 24 $\mu$m-selected ULIRGs (hereafter, 24 $\mu$m-ULIRGs) have also redshifts in the range $z=1-3$ \citep{Ho05,Ya05,Br07}.

The properties of these high-$z$ galaxies suggest that they are likely associated with an early phase in the formation of massive galaxies, supermassive black holes, and active galactic nuclei (AGNs). Therefore, making clear the origin of their energy sources for different redshift ranges is very important to understand the star formation history and the AGN activity in the high-$z$ Universe. 
In this context, a key point to be investigated is the relative AGN/starburst contribution to the bolometric 
luminosity of these galaxies. Here we present a quantitative estimate of the AGN content in submillimiter and 24~$\mu$m-ULIRGs, 
based on a deconvolution method recently developed by our group for local ULIRGs \citep[][hereafter N08]{Na08} by means of $Spitzer$-Infrared Spectrograph 
\citep[IRS;][]{Ho04} 
$5-8$~$\mu$m spectroscopy. 
Throughout this paper we use a standard cosmology 
with $H_{0}=73 \rm{km s^{-1} Mpc^{-1}}, \Omega_{\rm M}=0.3$ and $\Omega_{\Lambda}=0.7$.

%%%%%%%%%% Section 2: Sample Data and Data Reduction %%%%%%%%%%

\section[]{Sample Data and Data Reduction}

\begin{table*}
\centerline{\begin{tabular}{lcccc|lcccc}
\hline
submm-ULIRGs position & redshift & $L_{\rm IR}$ & ref. & P. ID  &24 $\mu$m-ULIRGs position & redshift & $L_{\rm IR}$ & ref. & P. ID \\
\hline
SMM J002634.10+170833.7 & 2.73 & 0.46     &  2 &3241   &  
SST24 J142538.22+351855.2 & 2.26 & 4.9 &  5 & 12  \\
SMM J023951.87-013558.8 & 2.81  & 2.29    &   2 & 3241    &  
SST24 J142626.49+344731.2 & 2.13 & 5.2   & 5 & 12\\
SMM J030227.73+000653.5 & 1.408 & 2.63    &     1 & 20081  &  
J142644.33+333052.0 & 3.355 & 2.39       & 7 & 12\\
SMM J094253.42+465954.5 & 2.38  & 5.79       &  2 & 3241  &  
SST24 J142645.71+351901.4 & 1.75 & 0.5   & 5& 12 \\
SMM J094303.69+470015.5 & 3.36  & 0.83    &   2 &  3241  &  
SST24 J142653.23+330220.7 & 1.86 & 0.8   & 5 & 12\\
SMM J105155.47+572312.7 & 2.67  & 0.46   &  2 &  3241    &  
SST24 J142804.12+332135.2 & 2.34 & 0.8  & 5  & 12\\
SMM J105207.49+571904.0 & 2.69  & 2.67    &  2 &  3241   &  
SST24 J142924.83+353320.3 & 2.73 & 1.5  & 5  &12 \\
SMM J105238.19+571651.1 & 1.852 & 0.55      &  1 & 20081 & 
SST24 J142958.33+322615.4 & 2.64 & 1.7  & 5 & 12 \\
SMM J123555.13+620901.6 & 1.875 & 0.6$^\dagger$ &  1 &20456& 
SST24 J143001.91+334538.4 & 2.46 & 4.2 & 5 &  12 \\
SMM J123600.16+621047.3 & 1.994 & 0.96$^\dagger$& 1 & 20456&  
J143028.52+343221.3 & 2.176 & 0.92     & 7  & 15 \\
SMM J123616.11+621513.5 & 2.578 & 0.66$^\dagger$&1 &20456 &  
SST24 J143251.82+333536.3 & 1.78 & 2.6  & 5 & 12 \\
SMM J123618.33+621550.4 & 1.865 & 0.64$^\dagger$&1 &20456 &  
J143312.7+342011.2 & 2.2 & 1.20      & 7 &   15  \\
SMM J123621.27+621708.1 & 1.988 & 0.69$^\dagger$& 1 &20456&  
SST24 J143358.00+332607.1 & 1.96 & 0.6   & 5 & 12\\
SMM J123634.51+621240.9 & 1.219 & 0.36$^\dagger$&  1 &20456 & 
J143424.24+334543.4 & 2.26 & 0.74     & 7 &  12  \\
SMM J123707.21+621408.1 & 2.484 & 1.23$^\dagger$& 1 &20081&  
SST24 J143447.70+330230.6 & 1.78 & 0.8  & 5 & 12 \\
SMM J123711.37+621331.1 & 1.996 & 0.5$^\dagger$ & 3 &20456& 
SST24 J143504.12+354743.2 & 2.13 & 0.8 & 5  & 12 \\
SMM J123711.98+621325.7 & 1.992 & 0.28$^\dagger$& 1 & 20081&
SST24 J143520.75+340418.2 & 2.08 & 0.9 & 5 &  12 \\
MM J154127.28+661617.0  & 2.79  & 0.9      &   2 & 3241   & 
SST24 J143523.99+330706.8 & 2.59 & 1.4 & 5 &  12 \\
SMM J163639.01+405635.9 & 1.495 & 0.61  &    1 &  20081   & 
SST24 J143539.34+334159.1 & 2.62 & 3.5  & 5 & 12 \\
SMM J163650.43+405734.5 & 2.378 & 3.10   &    1 & 20081   &  
SST24 J143644.22+350627.4 & 1.95 & 1.7 & 5 &  12 \\
SMM J163658.78+405728.1 & 1.21  & 0.09    &   2 &  3241  & 
IRS 9 J171350.001+585656.83 & 1.83 & 1.70& 6 & 3748 \\
SMM J163706.60+405314.0 & 2.38  & 0.64      &   2 & 3241  & 
IRS 11 J171439.570+585632.10 & 1.8 & 0.55& 6 & 3748\\
ERO J164502.26+462626.5 & 1.443 & $>$0.18    & 4 &13   & 
IRS 8 J171536.336+593614.76 & 2.6 & 1.89 & 6&3748 \\
SMM J221733.91+001352.1 & 2.555 & 0.77     &  1 & 20081   &  
IRS 2 J171538.182+592540.12 & 2.34 & 4.07& 6 &3748 \\ 
SMM J221735.15+001537.2 & 3.098 & 1.45     &  1 & 20081  &  
IRS 1 J171844.378+59200.53 & 2.1 & 3.03  & 6 &3748 \\
SMM J221735.84+001558.9 & 3.089 & 1.24     &  1 & 20081    & \\
SMM J221737.39+001025.1 & 2.614 & 2.02      & 1 & 20081    & \\
SMM J221804.42+002154.4 & 2.517 & 0.72     &  1 &  20081   & \\
\hline
\end{tabular}}
\caption{ 
Submm- and 24 $\mu$m-ULIRGs sample. Column 1: Object position. 
Col. 2, 3, and 4: Redshift and infrared luminosity,
in units of $10^{13}~L_\odot$ ($^\dagger$ \citep{Po08}) and references therein; 
[1] \citet{Ch05}; [2] \citet{Va07}; [3] \citet{Sw04}; [4] \citet{Gr05}; [5] \citet{Ho05}; 
[6] \citet{Ya05}; [7] \citet{Br07}. Col. 5: Programs of the $Spitzer$-IRS observations. \label{tb1}
}
\end{table*}
Our sample consists of all the submm- and 24~$\mu$m-selected sources with known redshift, and with an available $Spitzer$-IRS observation. 
The submm sample was obtained by combining the SCUBA detected sources from \citet{Ch05} (submm bright, pre-selected in radio) and \citet{Va07} (without radio pre-selection). These two subsamples are different in 
several respects (lensing properties, submm flux range, redshift distribution). However,
their selection criteria are not expected to introduce any bias in their AGN content (we discuss
possible effects in \S 4).
The 24~$\mu$m sources are also selected in different ways through their 24~$\mu$m flux and different optical/infrared colors (e.g.~24/8~$\mu$m , 24~$\mu$m/R). It is well known that these criteria are strongly biased toward sources whose mid-IR emission is dominated
by the AGN component. 
Redshift, bolometric luminosities as well as other relevant data for all the sources are presented in Table~1. 
The bolometric luminosities listed in Table~1 are obtained by fitting a spectral energy distribution to the available data
(see single references for details). The uncertainties on the single values can therefore be high (about 50\%
for submm sources, and a factor of three for 24~$\mu$m sources, since the mid-IR to bolometric correction is strongly
dependent on the AGN/SB relative contribution). 
Data reduction has been performed with the SPICE 1.4.1 package, 
using the pipeline version S15.3, provided by the {\em Spitzer} Science Center. 
The coadded images 
have been background-subtracted by means of the two observations in the nodding cycle. 
The calibrated one-dimensional spectra for the positive and the negative beams were extracted using the "optimal" 
extraction mode in SPICE, and the two one-dimensional spectra averaged in order to obtain the final spectrum.
Since we are interested in the average properties of the two populations, and the spectral quality is low, 
we stacked the spectra of the single 
sources, after shifting and re-sampling each spectrum to the rest-frame wavelengths and bin sizes. We also had a quick look at 
 each single spectrum, in order to check that the stacked data are not dominated by individual objects with high S/N.
\begin{figure*}
\includegraphics[width=18cm]{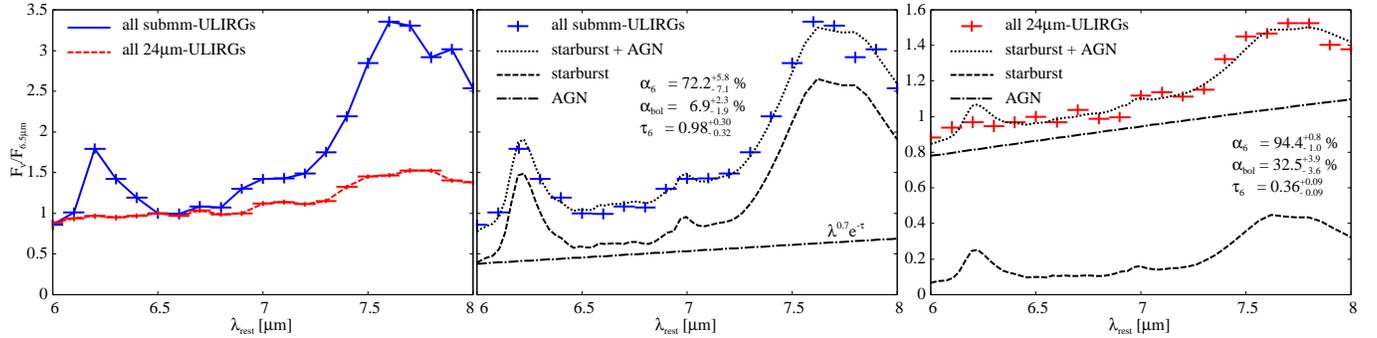}
\caption{$6 - 8~ \mu$m rest-frame composite spectra of submm-ULIRGs and 24~$\mu$m-ULIRGs, with their best fit models.
The flux density, $F_{\nu}$ are normalized at 6.5 $\mu$m. 
{\it Left}: Composite spectra for all submm-ULIRGs (solid blue line) and 24 $\micron$-ULIRGs (dashed red line). 
{\it Middle} and {\it right}: Composite spectrum of the submm-ULIRGs and 24 $\mu$m-ULIRGs, respectively, with the best fit model ({\it dotted line}) and the AGN ({\it dashed dotted line}) and starburst ({\it dashed line}) components.
$\alpha_{6}$ and $\alpha_{\rm bol}$ are the intrinsic AGN contribution to the 6 $\mu$m flux density 
(in percent) and the AGN contribution to the total infrared luminosity (in percent), respectively.
 $\tau_{6}$ is the optical depth to the AGN at 6 $\mu$m.}
\label{fig1}
\end{figure*}
\begin{figure}
\includegraphics[width=8.5cm]{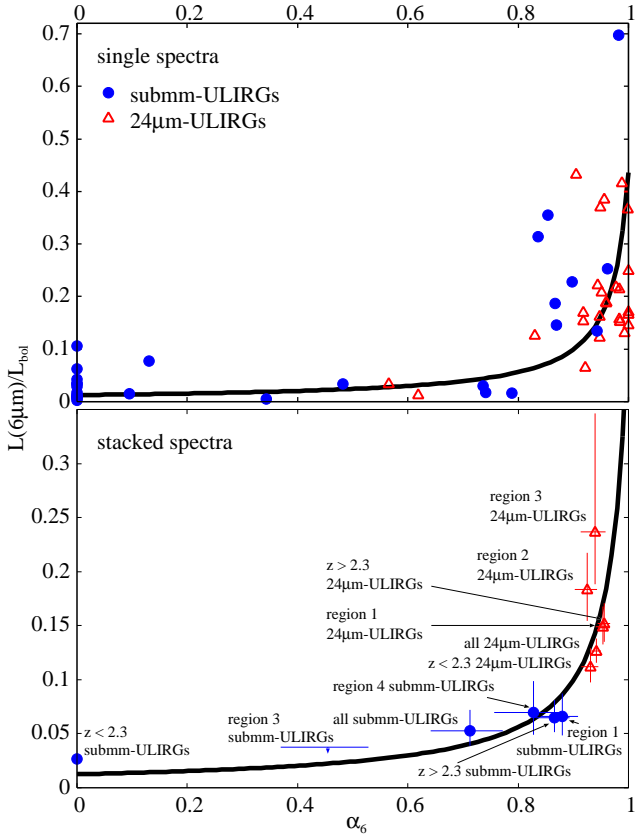}
\caption{
Upper panel: Absorption-corrected $6~\mu$m/bolometric ratio versus the AGN fraction, $\alpha_6$, for submm-ULIRGs ({\it blue filled 
circle}) and $24~\mu$m-ULIRGs sources ({\it red open 
triangle}). The continuous line refers to the best fit relation obtained for local ULIRGs (N08). Typical fit 
errors of x- and y- coordinates are a factor of 10\% and $2-3$, respectively.
Lower panel: same for the stacked spectra of the subgroups discussed in the text.}
\label{fig2}
\end{figure}

%%%%%%%%%% Section.3: AGN and Starburst Decomposition Model %%%%%%%%%%

\section{AGN and Starburst Decomposition Model}

In order to separate the AGN and the starburst components for submm-ULIRGs and 24 $\mu$m-ULIRGs, we use the spectral decomposition model  
for local ULIRGs of N08. The model is based on the following main points: 1) in ULIRGs, the ratio between the $3-8$~$\mu$m emission and the bolometric flux  
is $30-100$ times higher for AGNs than for starbursts. The contrast rapidly decreases with increasing wavelength.
2) The starburst spectrum in this wavelength range shows little variation from source to source, and a template
obtained by averaging the highest signal-to-noise pure starburst spectra fits well all the known starburst-dominated ULIRGs. 3) The AGN component can be represented with a
power law $f^{\rm AGN}_{\nu} \propto \lambda^{0.7}$ \citep{Ne07}, and an exponential attenuation $\exp({-\tau(\lambda)})$, where the optical depth follows the conventional law $\tau (\lambda) \propto \lambda^{-1.75}$ \citep{Dr89}. In this assumption, the total flux density is written as follows, 
\begin{equation}
f^{\rm obs}_{\nu}(\lambda) = \eta_{\rm SB} f^{\rm SB}_{\nu} + \eta_{\rm AGN} f^{\rm AGN}_{\nu} e^{-\tau(\lambda)},
\end{equation} 
where $\eta_{\rm SB}$ and $\eta_{\rm AGN}$ are the flux density amplitudes of the starburst and AGN templates $f^{\rm SB}_{\nu}$ and $f^{\rm AGN}_{\nu}$ normalized at 6 $\mu$m, respectively. 
We can estimate the only two free parameters (the ratio of $\eta_{\rm SB}$ to $\eta_{\rm AGN}$ and $\tau_{6}$, which is the 6 $\mu$m optical depth to the AGN) by fitting the spectrum.
From these, we can get the intrinsic AGN contribution to the 6 $\mu$m flux density, $\alpha_{6} = \eta_{\rm AGN} /  (\eta_{\rm AGN} + \eta_{\rm SB})$. In addition, we can also estimate the AGN contribution to the total infrared luminosity (this roughly corresponds to the bolometric luminosity, $L_{\rm bol}$, for ULIRGs), $\alpha_{\rm bol} = \eta_{\rm AGN} / (\eta_{\rm AGN} + K \eta_{\rm SB}) $, 
where $K = R^{\rm AGN}/R^{\rm SB}$, and $R^{\rm AGN}$ and $R^{\rm SB}$ are the ratios of the intrinsic 
flux at 6 $\mu$m 
to the total infrared flux of AGN and starbursts, respectively; in local ULIRGs, 
$\log R^{\rm AGN} = -0.36^{+0.06}_{-0.07}$, $\log R^{\rm SB} = -1.91^{+0.02}_{-0.02}$, yielding $K \sim 35$ (Nardini et al. in prep.). 
The model has been successfully applied to a sample of local ULIRGs, obtaining two important results: 1) despite the complexity of the  spectra, all the sources were successfully fitted,
showing that the relative AGN/starburst contribution and the extinction of the AGN component are 
responsible for most of the observed variety; 2) 
the expected AGN/starburst contributions to the bolometric luminosity reproduce closely the total 
observed luminosity. 
This is an important test for our model:
the bolometric contributions are
obtained from the 6~$\mu$m spectral decomposition only (and from the average bolometric ratios, which have a fixed value for all objects), 
therefore the comparison between the predicted and observed total luminosities is an independent check of the results. 

Here we apply the same model to high-redshift sources. In doing so, we assume that a) the 
intrinsic SED, and b) the bolometric ratios, are the same at low and high redshift. We discuss the implications and the limits of these assumptions in \S 4.

%%%%%%%%%% Section 4 Results %%%%%%%%%%

\section{Results: AGN/SB Deconvolution and redshift dependence of AGN activity}

The stacked spectra for the two samples of submm- and 24~$\mu$m-selected galaxies are shown in Fig.~1, together with our
deconvolution in the AGN and starburst components.
The main results are the following.\\
$\bullet$ With only two free parameters, the deconvolution model reproduces the observed average spectral emission of high redshift sources
in the 6-8~$\mu$m range, similarly to what happens for the low~z ULIRGs analized in N08. 
We note that the simple fit we applied is strongly based on the assumption of
little dispersion of the AGN and the starburst emission with respect to our templates. Therefore, it is not possible
to extend the method to longer wavelengths (considering the redshift of our sources, we have a good spectral
coverage up to $\sim12~\mu$m) due to the presence of extra features (mainly the silicate absorption) which are
well known to have quite different strengths from source to source.
\\
$\bullet$ We quantitatively estimate the intrinsic AGN contributions in submm-ULIRGs to the 6 $\mu$m and bolometric luminosities: 
$\alpha_{6} = 72.2^{+5.8}_{-7.1}$\%,  $\alpha_{\rm bol}=6.9^{+2.3}_{-1.9}$,  
where the errors refer
only to the formal statistical uncertainties in the
fits. \\
$\bullet$
For the 24~$\mu$m-ULIRGs we find $\alpha_{6}=94.4^{+0.8}_{-1.0}$\%
and  $\alpha_{\rm bol}=32.5^{+3.9}_{-3.6}$ \%.
It has already been suggested in previous studies \citep[e.g.,][]{Sa07}, and confirmed from the AGN and starburst $5-8$~$\mu$m to bolometric ratios (N08),
that $24~\mu$m-ULIRGs are dominated by the AGN component. The new result presented here is that 
the starburst component is nevertheless present, and dominates the total bolometric
luminosity.

The above results rely on a series of assumptions which need to be tested, in order to assess the
reliability of our method.
 In particular, 
as mentioned in \S 3, our choice of the $5-8$~$\mu$m rest-frame spectral interval is motivated by
(1) the constancy of the AGN and starburst spectra, and (2) the high AGN/SB flux ratio
in this band. These two properties are well measured for low-redshift sources, and the strongest
assumption of our work is that they are both valid at higher redshifts and luminosities. 
Moreover, the possible intrinsic scatter in AGN and SB templates introduces additional uncertainties
on top of the statistical errors in $\alpha_6$ and $\alpha_{\rm bol}$. A further uncertainty has to be added to
the estimate of $\alpha_{\rm bol}$ due to the possible intrinsic scatter in the AGN and the SB bolometric ratios.

Following N08, it is possible to test the method and give a realistic estimate of the uncertainties by comparing the measured
values of $\alpha_6$ with the intrinsic 6~$\mu$m/bolometric ratio
$R=L(6~\mu$m)/$L_{\rm bol}$. This new quantity has two interesting properties: 
1) it depends only mildly from our spectral modeling (only through the AGN absorption $\tau_6$) and never uses our assumptions on the bolometric ratios;
2) it is strongly sensitive
to the AGN content, being (at low redshift)
$\sim35$~times higher in AGNs than in starbursts.

We plot $R$ versus $\alpha_6$ in Fig.~2. The most relevant results  are: 
(1) the two quantities are correlated as expected,
i.e. the sources with the highest $\alpha_6$ also have the highest   $6~\mu$m/bolometric ratio;
(2) the continuous line in Fig.~2 represents the best-fit correlation obtained by N08
for local ULIRGs, and the overall good match between the curve and our points suggests that there are no
strong differences in bolometric ratios between low- and high- redshift sources.
The dispersion of the single points from the relation is of the order of a factor $\sim$2, while
the values for the composite spectra have a dispersion of $\sim$10-20\%. We consider these as the best estimates
of the errors involved in our analysis, and a confirmation of the validity of our approach.

\begin{table}
\label{tb2}
\centerline{\begin{tabular}{lccc|lccc}
\hline
Submm & $\alpha_{6}$ [\%]& $\alpha_{\rm bol}$ [\%]& 24-$\mu$m & $\alpha_{6}$ [\%]& $\alpha_{\rm bol}$ [\%]\\
\hline
all & $72.2^{+5.8}_{-7.1}$& $6.9^{+2.3}_{-1.9}$& all & $94.4^{+0.8}_{-1.0}$& $32.5^{+3.9}_{-3.6}$\\
$z<2.3$ & $0$& $0$& $z<2.3$ & $93.3^{+1.3}_{-1.5}$& $28.6^{+4.9}_{-4.4}$\\
$z>2.3$ & $87.2^{+2.6}_{-3.2}$& $16.3^{+3.7}_{-3.2}$& $z>2.3$ & $95.6^{+1.0}_{-1.2}$& $38.5^{+6.8}_{-5.9}$\\
Reg.1 & $88.5^{+2.8}_{-3.8}$& $18.1^{+5.1}_{-4.4}$& Reg.1 & $95.5^{+1.1}_{-1.3}$& $37.8^{+6.8}_{-5.9}$\\
Reg.3 & $45.4^{+15.2}_{-8.6}$& $2.3^{+1.9}_{-0.7}$& Reg.2 & $92.8^{+1.7}_{-2.1}$& $26.9^{+6.0}_{-5.1}$\\
Reg.4 & $83.5^{+5.0}_{-7.0}$& $12.7^{+5.4}_{-4.1}$ &  Reg.3 & $94.1^{+1.9}_{-2.5}$& $31.6^{+9.2}_{-7.6}$\\
\hline
\end{tabular}}
\caption{
Intrinsic AGN contribution to the 6~$\mu$m flux density, $\alpha_{6}$, and  
to the total infrared luminosity, $\alpha_{\rm bol}$ in each group of submm and 24~$\mu$m sources discussed in the text. See Fig.~4 for the definition of Reg.~1-4.
}
\end{table}

\begin{figure}
\includegraphics[width=8.0cm]{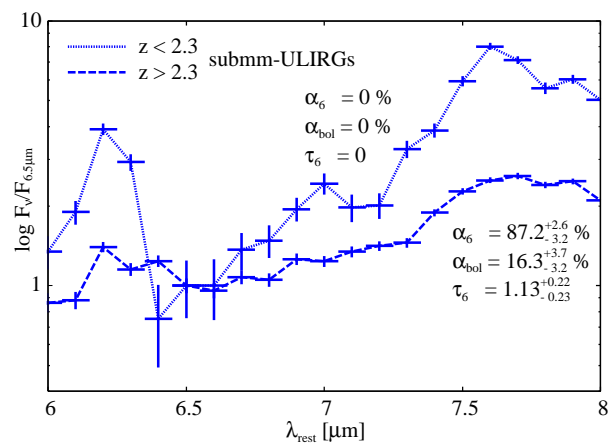}
\caption{Same as the left panel of Fig. \ref{fig1}, for submm-ULIRGs at $z$$<$2.3 ({\it dotted lines}) 
at $z$$>$2.3 ({\it dashed lines}.) The flux density, $F_{\nu}$ are normalized at 6.5 $\mu$m.}
\label{fig3}
\end{figure}

In order to further investigate the issue of the AGN/SB content of high redshift galaxies, we 
analyzed the composite spectra obtained from different redshift and infrared luminosity intervals.

In Fig.~3, we show the composite spectra for the submm sources at $z<2.3$ and $z>2.3$, respectively
($z = 2.3$, is the submm-ULIRG median redshift of \citet{Ch05}, and roughly divides our sample in
two subsamples of similar size). It is apparent from Fig.~3 that the polycyclic aromatic hydrocarbon (PAH) emission features of lower-$z$ ($z$$<$2.3) submm-ULIRGs are stronger than those of higher-$z$ ($z$$>$2.3) 
galaxies. The
AGN contribution is 
not needed in the fit of the low-z sample, while 
 $(\alpha_{6}[\%], \alpha_{\rm bol}[\%]) =(87.2^{+2.6}_{-3.2}$, $16.3^{+3.7}_{-3.2}$)  for the high-$z$ sample. 

\begin{figure}
\includegraphics[width=8.5cm]{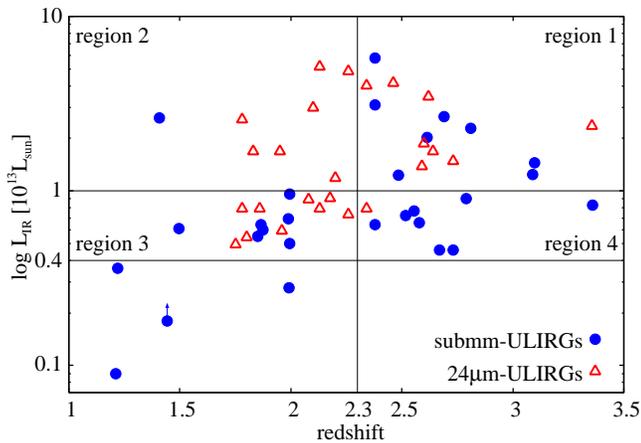}
\caption{Luminosity--redshift plot for our samples of submm-ULIRGs ({\it blue filled 
circle}) and 24 $\mu$m-ULIRGs ({\it red open triangle}). We define four regions
delimited by the boundaries at $z=2.3$,  
$L_{\rm IR} = 0.4\times 10^{13}~L_{\odot}$ and $L_{\rm IR} = 10^{13} L_{\odot}$.}
\label{fig4}
\end{figure}

\begin{figure*}
\includegraphics[width=18cm]{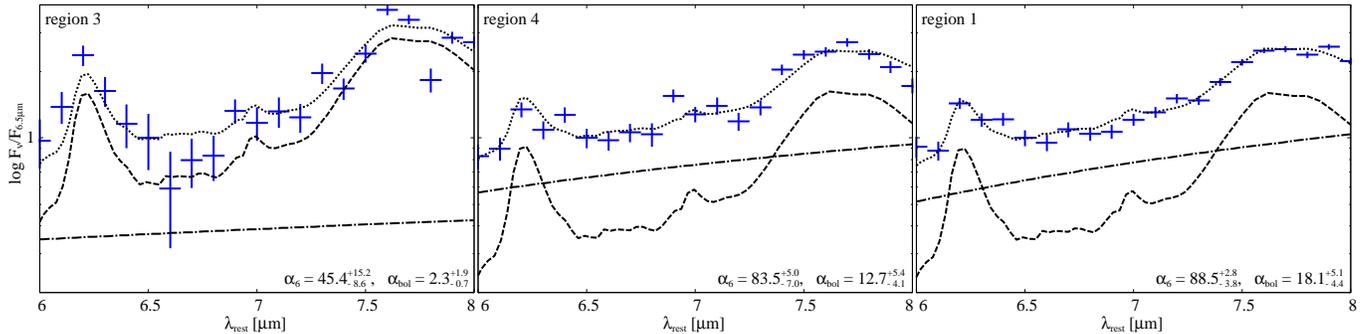}
\caption{Same as the middle panel of Fig. \ref{fig1}, for different subgroups of submm-ULIRGs:
low-$z$, low luminosity sources ({\it left}), high-$z$, low luminosity sources ({\it middle}), and
high-$z$, high luminosity sources (right).}
\label{fig5}
\end{figure*}

This correlation needs to be checked for several selection/physical effects which could affect its significance and interpretation.
In particular, we considered possible effects due to (1) selection, (2) K-correction , and (3) luminosity:\\
1) Selection effects. The submm galaxies in our sample are obtained from \citet{Ch05} and \citet{Va07}.
The different selection criteria, based on radio pre-selection, lensing properties and submm flux are not expected to
be relevant for our results. However, we note that almost all the \citet{Va07} objects have z$>$2.3 (Table~1), while
those of \citet{Ch05} span the whole redshift range. In order to check whether the redshift effects in Fig.~3 are related to
the properties of the two different samples, we repeated our analysis for the \citet{Ch05} sample only, obtaining
results perfectly compatible with those in Fig.~3. In other words, no significant difference is found between the \citet{Va07}
and \citet{Ch05} objects at z$>$2.3.\\
2) K-correction effects. The SCUBA $850~\mu$m selection does not seem to introduce any particular
bias in favor of one of the two classes of sources: moving from redshift $z=1.8$ to $z=2.7$ 
(the median redshifts for the two subsamples of submm-ULIRGs),
the observed central wavelength changes from $\lambda=300~\mu$m to $\lambda=230~\mu$m.
Indeed, we checked that there is no difference between the AGN and starburst bolometric ratios at these two wavelengths by using the SEDs of the ULIRGs classified as AGNs and starburst, respectively \citep{Ya07}. \\
3) Luminosity effects.
Since the AGN fraction in local ULIRGs increases with increasing infrared luminosity \citep{Ve97,Ki98}, 
our results may imply that the real physical trend is with luminosity, which is obviously partially degenerate with redshift. In order to check for this possibility, we analyze separately the luminosity and redshift dependence. 
In Fig.~4 we show the luminosity-redshift plot of our sample. 
Analyzing the population of the four highlighted regions, we note that we can disentangle
the luminosity and redshift effects by building composite spectra for submm-ULIRGs in regions 1 ($L>10^{13}~L_\odot, z>2.3$), 
3 ($L=0.4-1\times10^{13}~L_\odot, z<2.3$) and 4 ($L=0.4-1\times10^{13}~L_\odot, z>2.3$).
The results of the fits of these templates are presented in Table~2.
Objects in regions 1 and 4 are in the same redshift interval but have quite different luminosities,
while sources in regions 3 and 4 have similar luminosities but different redshifts.

The spectra and the results are shown in Fig.~5. 
For submm-ULIRGs, the redshift dependence 
is confirmed by the comparison between the average spectra from the same luminosity range,
while no strong luminosity dependence is found by comparing the composite spectra at $z>2.3$.
The increase of the AGN fraction with the redshift, and its constancy with the luminosity
are also indicated by the average $6~\mu$m/bolometric ratios, plotted in Fig.~2
as a function of the average AGN contributions. 

We have repeated the analysis the $L-z$ plane also for the 24~$\mu$m sources, building compositte
spectra for Reg.~1,~2 and 3. The fit results presented in Table~2 indicate that no luminosity or redshift effect can be significantly detected.

%%%%%%%%%% Section 5: Conclusions %%%%%%%%%%

\section{Conclusions}

We quantitatively estimated the relative AGN and starburst contributions in two samples of
submilliter- and 24 $\mu$m-selected galaxies in the redshift range $z=1-3$ by applying a
deconvolution model based on AGN and starburst templates, already tested on low-redshift sources 
to the  
the stacked $6-8$~$\mu$m spectra.
We showed that the method gives consistent
results on high redshift sources, suggesting that both the average $6-8$~$\mu$m spectra
and the 6~$\mu$m/bolometric ratios are similar in local and high redshift  ULIRGs.
Overall, we find that the starburst component is dominant in submm-ULIRGs both at 6~$\mu$m and bolometrically. We find evidence of a higher AGN contribution in higher redshift ($z>2.3$) submm-ULIRGs with respect to lower redshift ($1<z<2.3$) objects.
The AGN component always dominates the mid-infrared emission in 24~$\mu$m-ULIRGs, as already known from previous work. However, the starburst contribution to the total luminosity 
is of the same order, or slightly higher, than that of the AGN even in the 24~$\mu$m-ULIRGs.

\section*{Acknowledgments}
We thank the anonymous reviewer for valuable comments. This work has been partly supported by grant PRIN-MUR 2006025203.

\end{document}